\documentclass[a4paper]{article}

\usepackage[english]{babel} 
\usepackage[small]{caption}
\usepackage{url}
\usepackage{algorithm}
\usepackage{algorithmic}
\usepackage{placeins}
\usepackage{graphicx} 

\usepackage{amsthm}

\theoremstyle{definition}
\newtheorem{mydef}{Definition}

\def\addsymbol #1: #2#3{$#1$ \> \parbox{5in}{#2 \dotfill \pageref{#3}}\\}

\begin{document}

\title{Comparison of Bucket Sort and RADIX Sort}
\author{Panu Horsmalahti\\
\texttt{panu.horsmalahti@tut.fi}}
\maketitle





\begin{abstract}
Bucket sort and RADIX sort are two well-known integer sorting algorithms. This paper measures empirically what is the time usage and memory consumption for different kinds of input sequences. The algorithms are compared both from a theoretical standpoint but also on how well they do in six different use cases using randomized sequences of numbers. The measurements provide data on how good they are in different real-life situations. 

It was found that bucket sort was faster than RADIX sort, but that bucket sort uses more memory in most cases. The sorting algorithms performed faster with smaller integers. The RADIX sort was not quicker with already sorted inputs, but the bucket sort was.
\end{abstract}

\pagenumbering{arabic}

\section{Introduction}
\label{sec-introduction}

Sorting algorithms have a long history and countless of algorithms have been devised~\cite[p. 2]{gpu}, not only because of their real-life applications. Most programs use sorting and sorting algorithms should therefore be as fast as possible. One commonplace applicaton is for databases which may need to sort tens of millions of objects~\cite[p. 2]{database}. Making a faster sorting algorithm for a particular use case can save significant amounts of money. It is important to have good empirical data on sorting algorithms to choose the best sorting algorithm for each use case. No single sorting algorithm is best for all use cases~\cite[p. 1]{bestsorting}. The main factors in choosing the algorithm are time usage, memory consumption and ease of implementation.

This paper compares two well known sorting algorithms, \emph{bucket sort} and \emph{RADIX sort}. Bucket sort and RADIX sort are integer sorts, which are used to sort a sequence of integers instead of the more general comparison sorts. Related studies by Curtis~\cite{bestsorting} and by Loeser~\cite{loeser} have focused on comparison sorting algorithms. Integer sorting algorithms can also be used to sort text strings for example, since each string can be converted into an integer.

Bucket sort operates by creating a number of buckets. Each integer is placed into a bucket based on the integer value. A common way is to divide the input element by the integer range, thus defining an integer range for each of the bucket. All buckets are then sorted. After that, the buckets are concatenated into a single list, which is the output of the algorithm.~\cite[p. 8]{popular}

RADIX is an integer sort which sorts all elements using a single digit at a time. This sorting algorithm was used by very old card-sorting machines~\cite[p. 150]{algorithms}. The array is first sorted on the basis of the largest digit. After that, all cards are sorted using the second largest digit and so on. While it is intuitive to sort using the \emph{MSD} (most significant digit) first, it is actually more efficient to sort with respect to the least significant digit. Although the \emph{LSD} (least significant digit) RADIX sort is unintuitive, it works too. This paper uses the LSD version of the RADIX sort.

Theoretical time complexities do not tell much about how fast sorting algorithms are in real-life applications. The input array does not often have a uniform distribution, the integer range may be very small or really large and sometimes the array is already sorted. This paper will empirically measure the time usage and memory consumption on these inputs and will also compare them to a uniform distribution of integer values. Each test sorts up to 100 million keys.

This paper answers the problem of how well these sorting algorithms do in real-life use cases. Input sequences can be biased in numerous ways, therefore six different cases are selected for the benchmark along with a uniform distribution of random integers which is used as a control. The algorithms are benchmarked with increasing size of the input data, and the growth rate of the time and the absolute time for the largest input will tell which is better for the selected use cases.

The results show that bucket sort is faster in all test cases, but it also uses more memory than the RADIX sort in some cases. Bucket sort is slow with large integer rages. RADIX sort is equally as fast for sorted inputs as it is for unsorted inputs. The time usage increases linearly with respect to the size of the input as is predicted by theory.

The paper is organized into four sections. After introduction, the Section~\ref{sec-theory} briefly describes sorting algorithms theory, and how integer sorting differs from comparison sorting. Both algorithms are also described in detail. Section~\ref{sec-comparison} describes what input arrays are chosen and how they are generated. The testing procedure is explained in detail. Finally, the Section~\ref{sec-results} describes the results and how they can help in choosing the correct sorting algorithm.

\section{Sorting Algorithm Theory}
\label{sec-theory}

A Sorting algorithm takes as an input a sequence of elements, and outputs a permutation of the sequence. Additionally, all elements in the output sequence must be in (e.g. nondecreasing) order, using some come comparison function.  General sorting algorithms, like merge sort, are neutral with respect to the data type. In essence they work for all kinds of data which can be compared. 

\subsection{Notation}

Asymptotic notations are used in asymptotic analysis to describe how an algorithm behaves with respect to input size. They can describe how much memory or time consumption increases when the input size is increased. Usually we are interested in only the asymptotic growth rate, which means that we are only interested in the largest term and not constants factors. 

The \emph{input size} $n$ depends on the problem to be studied. It can be the number of items in the input array, the number of bits needed to represent the input, or it can even be two different numbers if the input would be a graph. The \emph{running time} $f(n)$ is the number of operations or steps executed. The time required to execute each line in pseudocode is constant. In real life, different computers execute different operations using different amounts of time, but when $n$ grows large enough, these differences become insignificant.~\cite[p. 25]{algorithms}

Definitions \ref{onotation}, \ref{omeganotation} and \ref{thetanotation} are used to describe the growth rate of an algorithm. Note that some publications use set theory notation instead of the equality sign.

\begin{mydef}
\label{onotation}
We say that $f(n) = O(g(n))$, if there exist constant $c > 0$ and $N$ such that $f(n) \leq cg(n)$ for all $n \ge N$~\cite[p. 25]{efficient}
\end{mydef}

\noindent
In practical terms the $O$-notation in Definition \ref{onotation} describes the worst-case behaviour of the algorithm. It guarantees that the longest time the algorithm can use is less than or equal to $g(n)$ as $n \to \infty$. For example, if the growth rate for the worst case would be $f(n) = 2n^2$, we can say $f(n) = O(n^2)$.

\begin{mydef}
\label{omeganotation}
We say that $f(n) = \Omega(g(n))$, if there exist constant $c, N$ such that $f(n) \ge cg(n)$ for all $n \ge N$~\cite[p. 25]{efficient}
\end{mydef}

\noindent
The $\Omega$-notation in Definition \ref{omeganotation} tells the best-case behaviour of the algorithm. It tells us the minimum running time of the algorithm.

\begin{mydef}
\label{thetanotation}
We say that $f(n) = \Theta(g(n))$, if $f(n)=O(g(n))$ and $g(n)=O(f(n))$~\cite[p. 25]{efficient}
\end{mydef}

\noindent
The $\Theta$-notation in Definition \ref{thetanotation} describes both the worst-case and best-case growth rates for the algorithm.

It has been proven that comparison-based sorting algorithms have a lower bound of $O(n\log n)$, where $n$ is size of the input~\cite[p. 146]{algorithms}. However, lower bound does not apply to integer sorting, which is a special case of sorting~\cite[p. 49]{popular}. Integer sorts can be faster than comparison based sorting algorithms because they make assumptions about the input array. Both the RADIX sort and the bucket sort are integer sorting algorithms.

A sorting algorithm is \emph{stable}, if the order of equal elements in the input array remains the same in the output array~\cite[p. 149]{algorithms}. In the case of RADIX, stableness depends on the underlying digit sorting algorithm~\cite[p. 150]{algorithms}. Bucket sort is also stable, if the underlying sorting algorithm is stable.

The time complexities of bucket sort and RADIX sort are well known, but they vary depending on which variant of the sort is used. Normal bucket sort has time complexity of $\Theta(n+r)$ where r is the range of numbers~\cite[p. 155]{algorithms}. RADIX sort has a time complexity of $\Theta(d(n+k))$~\cite[p. 151]{algorithms}, where $d$ is the number of digits in the largest integer and each digit can take $k$ values. The average time complexity is the same as the worst case for both algorithms.

The algorithms in this paper are written in pseudocode notation. This pseudocode notation is easy to translate into a real programming language. Assignment is written using $\gets$. Loops begin with {\bf for} or {\bf while} and are closed with {\bf end}. The to-clause in the {\bf for} loop is indicated with $\to$. $length[A]$ refers to the length of the array $A$. Sometimes operations are described in plain english, e.g. \texttt{stable sort array A}, as any stable sort is suitable for the algorithm.

\subsection{Insertion sort}
\label{sec-insertion}

\emph{Insertion sort} works in a similar way to a human sorting a few playing card one at a time. Initially, all cards are in the unsorted pile, and they are put one by one to the left hand in the correct position. The sorted list of cards grows until all cards are sorted.~\cite[p. 18]{algorithms}

Sorting is usually done in-place. The algorithm goes through the array $k$ times, and in each iteration places the number A[j] in the correct position of the already sorted array. It is efficient for small inputs which is why it is chosen for the bucket sort implementation. It is also simple to implement. 

Insertion sort is stable~\cite[p. 44]{popular}, as all equal elements are inserted after the last equal one in the sorted array. This is required for the bucket sort to be stable. In the worst case the sorted section is completely shifted in every iteration, resulting in $O(n^2)$.  In the best case the algorithm is $\Omega(n)$, when input array is already sorted. The pseudocode in Algorithm \ref{insertionsort} assumes that indexing starts from zero.~\cite[p. 44]{popular}

\begin{figure}[t]
\begin{algorithm}[H]
\caption{INSERTION-SORT(A)}
\begin{algorithmic}
\FOR{$j \gets 1 \to length[A]-1$}
\STATE $key \gets A[j]$
\STATE $i \gets j-1$
\WHILE{$i \geq 0 \land A[i] > key$}
\STATE $A[i+1] \gets A[i]$
\STATE $i \gets i-1$
\ENDWHILE
\STATE $A[i+1] \gets key$
\ENDFOR
\end{algorithmic}
\end{algorithm}
\caption{Pseudocode of the insertion sort algorithm \cite[p. 44]{popular}}
\label{insertionsort}
\end{figure}


\subsection{Bucket sort}

This paper uses the generic form of bucket sort. It is assumed that each integer is between $0$ and $M$. $B[1\ldots n]$ is an array of buckets (for a total number of $n$ buckets) which in this implementation are linked lists. Each input element is inserted into a bucket $B[n\cdot A[i]/M]$. They are then sorted with the insertion sort, which is decribed in Section \ref{sec-insertion}. A pseudo-code version of bucket sort~\cite[p. 153]{algorithms} is shown in Algorithm \ref{bucketalgo}.

\begin{figure}[t]
\begin{algorithm}[H]
\caption{BUCKET-SORT(A)}
\label{bucketalgo}
\begin{algorithmic}
\STATE $n\gets length[A]$
\FOR{$i \gets 1 \to n$} 
\STATE insert $A[i]$ into list $B[n\cdot A[i]/M]$
\ENDFOR
\FOR{$i \gets 0 \to n-1$}
\STATE sort list $B[i]$ with insertion sort
\ENDFOR
\STATE 	concatenate lists $B[0]$,$B[1]$, $\ldots$, $B[n-1]$ together

\end{algorithmic}
\end{algorithm}
\caption{Pseudocode of the bucket sort algorithm \cite[p. 153]{algorithms}}
\end{figure}

\FloatBarrier

\noindent
The sorting algorithm assumes that the integers to be sorted tend to have an uniform distribution, which is the key to the performance of this algorithm. For example, if $n$ integers are sorted exactly into $n$ buckets, the running time is $\Theta (n)$. If the integers are not uniformly distributed, the algorithm may still run in linear time if the sum of the squares of the bucket sizes is linear in the total number of elements~\cite[p. 155]{algorithms}. 

The more uneven the input distribution is, the more the algorithm slows down, since more elements are put into the same bucket. Bucket sort is stable, if the underlying sort is also stable, as equal keys are inserted in order to each bucket.

\subsection{Counting sort}
\label{sec-counting}

\emph{Counting sort} works by determining how many integers are behind each integer in the input array $A$. Using this information, the input integer can be directly placed in the output array $B$. Counting sort is stable~\cite[p. 149]{algorithms}, which is important as it is used in the RADIX sort.

All numbers are assumed to be between $0$ and $k$. In the pseudocode in Algorith \ref{countingalgo}, $C[i]$ first holds the number of input integers equal to $i$ after the second for-loop. Then $C[i]$ is modified to hold the number of integers less than or equal to $i$, which can be used to place the integers. In the final loop, integers are directly placed in the correct position to the output array $B$. $C[A[j]]$ is decremented so that the next $A[j]$ is placed one position to the left.
~\cite[p. 29]{algorithms}

\begin{figure}[t]
\begin{algorithm}[H]
\caption{COUNTING-SORT(A, B, k)}
\begin{algorithmic}

\FOR{$i \gets 0 \to k$}
\STATE $C[i] \gets 0$
\ENDFOR

\FOR{$j \gets 1 \to length[A]$}
\STATE $C[A[j]] \gets C[A[j]]+1$
\ENDFOR

\FOR{$i \gets 1 \to k$}
\STATE $C[i] \gets C[i] + C[i-1]$
\ENDFOR

\FOR{$j \gets length[A] \to 1$}
\STATE $B[C[A[j]]] \gets A[j]$
\STATE $C[A[j]] \gets C[A[j]]-1$
\ENDFOR
\end{algorithmic}
\end{algorithm}
\caption{Pseudo code of the counting sort algorithm \cite[p. 148]{algorithms}}
\label{countingalgo}
\end{figure}

\subsection{RADIX sort}

RADIX is a sorting algorithm which sorts all elements on the basis of a single digit at a time. RADIX sort is not limited to sorting integers, because integers can represent strings. There are two main variations of the RADIX. The first one starts from the most significant digit (MSD) and the second from the least significant digit (LSD). RADIX sort is also useful when sorting records with multiple fields, like year, month and day. RADIX sort could first sort it on the day, then the month and finally the year.~\cite[p. 150]{algorithms}

First, the array is sorted using the LSD. For each pass a algorithm sorts it by a digit. This paper uses the \emph{counting sort}, described in Section \ref{sec-counting}, because it is efficient if the integer range is small (e.g. $0\ldots 9$) and it is also stable. Next step is to sort it using the second LSD and so forth. The last step sorts it using the 
MSD. In the end, all the elements are sorted. 

RADIX sort is stable, as the chosen underlying digit sorting algorithm is stable.~\cite[p. 150]{algorithms}. If the input integers have at most $d$-digits, then the algorithm will go through the array $d$ times, once for each digit. The pseudo-code for RADIX sort is shown in Algorithm \ref{radixalgo}.

\begin{figure}[t]
\begin{algorithm}[H]
\caption{RADIX-SORT(A, d)}
\begin{algorithmic}
\FOR{$i \gets 1 \to d$} 
\STATE stable sort array $A$ on digit $i$
\ENDFOR
\end{algorithmic}
\end{algorithm}
\caption{Pseudo code of RADIX sort algorithm \cite[p. 151]{algorithms}}
\label{radixalgo}
\end{figure}

In the pseudocode the array to be sorted is $A$ and the number of digits in the largest integer is $d$. If we use the counting sort each pass will use $\Theta(n+k)$ time, where each digit is in the range of $0\ldots k-1$. The whole sort takes $d$ passes, so the total time usage is $\Theta(d(n+k))$.~\cite[p. 151]{algorithms}.

\section{Comparison of the Algorithms}
\label{sec-comparison}

The sorting algorithms are now compared empirically by measuring time usage and memory consumption. To measure the sorting algorithm following inputs are used, where $n$ is the number of elements to be sorted. All input cases are measured using three different sizes: $n=10^6$, $n=10^7$ and $n=10^8$. All input integers are between $0$ and $M$.

\subsection{Test cases and implementation}

Six different input cases and three different input sizes are empirically measured to find out how the algorithms perform.

\begin{enumerate}
\item $n$ integers evenly distributed and in random order, $M=10^6$
\item $n$ integers already sorted, $M=10^6$
\item $n$ integers of which 95\% already sorted , $M=10^6$
\item $n$ integers with small range of $M=10^4$
\item $n$ integers with large range of $M=10^8$ 
\item $\frac{n}{3}$ of the integers with the same value $k$, rest of the integers all with different values, $M=10^6$
\end{enumerate}

The above inputs show how the two algorithms work in different real-life use cases. Time usage and memory consumption is measured. The first input is used as a control, as sorting algorithms often assume that the data  is uniformly distributed. Sorting algorithms are often analyzed and tested using a uniform distribution~\cite[p. 1]{algorithms}. The second input is important to measure, since often the data is already sorted and the algorithm should still perform well. 

The third input array is a common use case, since the inputs are often almost sorted. The fourth checks how well the algorithm works for a large number of integers with a small range (e.g. many values might be the same).

The fifth test is for a large range, which is tested since some integer sort implementations are slow with large integer ranges. The sixth and final case is to check how well the algorithm copes with a large number of the same value. Bucket sort is expected to suffer from this use case a lot.

The algorithms were implemented using the C++ programming language. The same C++ program also created the inputs. Random integers are created using \texttt{rand()} function. The seed number was not randomized, so that each test could be run multiple times with the same input. 

The implementation uses the vector$<$int$>$ container. Time usage is measured using \texttt{clock()} function. When sorting the same input array using the same algorithm multiple times, the time usage varied less than 5\%, so time measurement is assumed to be sufficiently accurate. 

Memory consumption is measured using the task manager. A standard laptop with a Intel i5 processor was used for the benchmark using the GNU/Linux operating system.
The results are shown in Table \ref{timetable} and Table \ref{memtable}.

\subsection{Results}

\FloatBarrier

\begin{table}[t]                       
\begin{center}                          
\caption{Measured time consumptions [s]}       
\label{timetable}
\begin{tabular}{r|ccc|ccc}               
 & \multicolumn{3}{|c|}{RADIX sort} & \multicolumn{3}{|c}{Bucket sort}  \\
Input no. &  $n=10^6$ & $n=10^7$ & $n=10^8$ & $n=10^6$ & $n=10^7$ & $n=10^8$  \\
\hline
1 & 1.78  & 17.66  & 176.84 & 0.74 & 5.84 & 45.92  \\
2 & 1.88 & 18.93 & 174.85 & 0.61 & 4.14 & 26.83\\
3 & 1.86 & 17.69 &  178.68 & 0.61 & 4.12 & 28.27 \\
4 & 1.05 & 11.10 & 105.31 & 0.31 & 3.00 & 31.54 \\
5 & 2.45 & 24.56 & 246.28 & 0.78 & 8.54 & 97.43 \\
6 & 1.77 & 17.66 & 176.65 & 0.61 & 5.18 & 41.06 \\

\end{tabular}
\end{center}
\label{table:taulukko1}
\end{table}

\begin{table}[t]                       
\label{table:taulukko2}
\begin{center}                          
\caption{Measured memory usage [MB]}       
\label{memtable}
\begin{tabular}{r|r@{.}lr@{.}lr@{.}l|r@{.}lr@{.}lr@{.}l}    
 & \multicolumn{6}{|c|}{RADIX sort} & \multicolumn{6}{|c}{Bucket sort}  \\
Input no. &  \multicolumn{2}{c}{$n=10^6$} & \multicolumn{2}{c}{$n=10^7$} & \multicolumn{2}{c|}{$n=10^8$}
& \multicolumn{2}{c}{$n=10^6$} & \multicolumn{2}{c}{$n=10^7$} & \multicolumn{2}{c}{$n=10^8$} \\
\hline
1 & 5&1  & 39&4 & 382&7 & 24&4 & 117&4 & 382&8 \\
2 & 20&6  & 79&2 & 382&9 & 47&2 & 117&4 & 382&8\\\
3 & 19&5 & 76&7 & 382&9 & 24&4 & 109&4 & 382&8\\
4 & 5&1 & 39&4 & 382&7 & 31&9 & 39&5 & 382&8 \\
5 & 5&1 & 39&4 & 382&7 & 24&4 & 232&8  & 2344&0\\
6 & 5&1 & 39&4 & 382&7 & 19&9 & 96&7 & 382&8\\
\end{tabular}
\end{center}

\end{table}

\begin{figure}[t]
  \centering
      \includegraphics[scale=0.53]{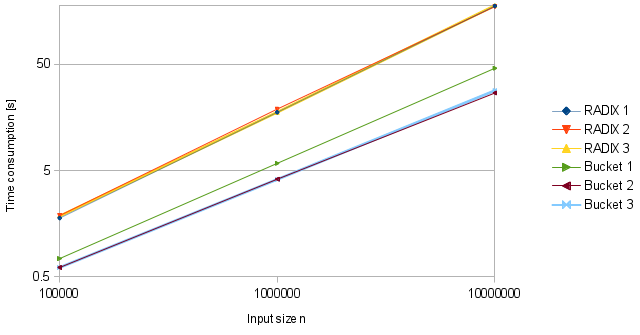}
  \caption{Time consumption for inputs 1, 2 and 3}
  \label{kaavio1}
\end{figure}

\begin{figure}[t]
  \centering
      \includegraphics[scale=0.53]{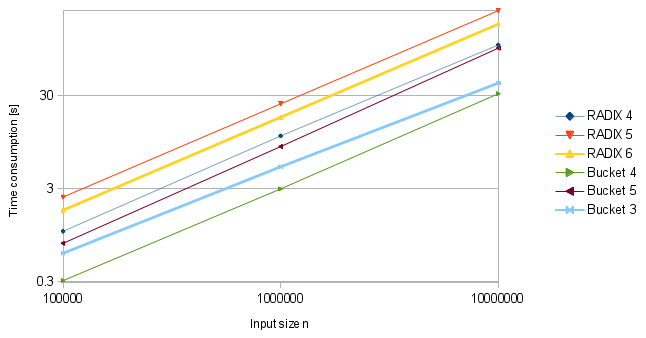}
  \caption{Time consumption for inputs 4, 5 and 6}
  \label{kaavio2}
\end{figure}

Table \ref{timetable} summarises the time consumption and Table \ref{memtable} summarises the memory consumption. Figure \ref{kaavio1} shows a graph for input cases 1, 2 and 3 and the Figure \ref{kaavio2} for inputs 4, 5 and 6. The diagrams are in logarithmic scale. The time usage increases linearly according to the results as is predicted by theory.

On the whole, bucket sort is faster in all cases, but uses significantly more memory, except when $n = 10^8$. In the fifth input case the bucket sort uses an order of magnitude more memory than RADIX. The reason why bucket sort is faster in all cases may be implementation specific. A further study with a different implementation might clarify whether or not the reason is implementation specific.

The results for inputs no. 2 and 3 indicate that RADIX sort performs as well for unsorted and sorted inputs. Bucket sort on the other hand is clearly faster for sorted arrays. The reason might be that the underlying insertion sort is fast for sorted input arrays. The results for the fourth input sequence indicates that RADIX sort performs well with smaller integers, because then $d$ is smaller. Bucket sort is also faster than the default case.

Both algorithms are quite slow in the fifth input array with the large range of $M = 10^8$, and RADIX is slow because of the higher number of digits $d$. This is the slowest input array for both of the algorithms.

RADIX performs just as fast for the sixth input sequence as it does for the control sequence with uniform distribution. The same is true for bucket sort, even though bucket sort has to use the insertion sort for third of the numbers.

\FloatBarrier

\section{Summary}
\label{sec-results}

Two sorting algorithms were compared both empirically and theoretically. Six different use cases were identified and measurements were made using three different input sizes, ranging three orders of magnitude. The first input had a uniform distribution of random numbers. The second input was sorted, which tested how well the algorithms perform with fully sorted input sequences.

The third input had 95\% of numbers sorted, which tested for nearly sorted input arrays. The fourth and fifth inputs had a small range and a large range, respectively, to test how the algorithm react. The last case tested an input with a large amount of numbers with a same value.

RADIX sort used the least significant digit version and the counting sort. Bucket sort used the insertion sort as the underlying sorting algorithm.

The algorithms and the creation of input arrays were implemented using the C++ programming language. Time consumption and memory usage were empirically measured. The sorting took up to a hundred seconds with the largest input.

It was found out that bucket sort is faster in all cases. The performance of the RADIX sort is slow only when the range of the integers is rather large. The bucket sort was found also to be slow with large integer ranges. The bucket sort was found to be quite fast with small integer ranges, which is also true for the RADIX sort. RADIX sort is as quick for unsorted inputs as it is for sorted inputs. The memory usage of the RADIX sort is slightly better than the bucket sort when sorting a small number of integers. Bucket sort uses large amounts of memory when sorting numbers with a large range.

More research should be made in the future to compare these sorting algorithms with comparison based ones like the merge sort. Studies should also be made using parallel versions of bucket sort and RADIX sort.


\addcontentsline{toc}{chapter}{References} 
\renewcommand{\bibname}{References} 
\bibliographystyle{plain}
\bibliography{lahteet}

\begin{thebibliography}{1}

\bibitem{efficient}
Eric Bach and Jeffrey Shallit.
\newblock {\em ``{A}lgorithmic {N}umber {T}heory: {E}fficient {A}lgorithms"},
  volume~1 of {\em Foundations of Computing}.
\newblock August 1996.

\bibitem{popular}
C.~Canaan, M.~S. Garai, and M.~Daya:.
\newblock ``{P}opular sorting algorithms''.
\newblock {\em World Applied Programming}, 1(1):42--50, April 2011.

\bibitem{bestsorting}
Curtis~R. Cook and Do~Jin Kim.
\newblock ``{B}est sorting algorithm for nearly sorted lists".
\newblock {\em Commun. ACM}, 23(11):620--624, November 1980.

\bibitem{algorithms}
T.~H. Cormen, C.~E. Leiserson, R.L Rivest, and C.~Stein.
\newblock {\em ``{I}ntroduction to {A}lgorithms"}.
\newblock MIT Press, 2nd edition edition, August 2001.

\bibitem{database}
G.~Graefe.
\newblock ``{I}mplementing sorting in database systems".
\newblock {\em ACM Comput. Surv.}, 38, September 2006.

\bibitem{loeser}
Rudolf Loeser.
\newblock ``{S}ome performance tests of “quicksort” and descendants".
\newblock {\em Commun. ACM}, 17(3):143--152, March 1974.

\bibitem{gpu}
N.~Satish, M.~Harris, and M.~Garland.
\newblock ``{D}esigning efficient sorting algorithms for manycore gpus".
\newblock In {\em Parallel \& Distributed Processing}, pages 1--10, May 2009.

\end{thebibliography}
\end{document}